\documentclass[12pt,preprint]{aastex}

\def\simgt{\lower 2pt \hbox{$\, \buildrel {\scriptstyle >}\over{\scriptstyle \sim}\,$}}
\def\simlt{\lower 2pt \hbox{$\, \buildrel {\scriptstyle <}\over{\scriptstyle \sim}\,$}}

\begin{document}

\title{{\it Chandra} observations of the hybrid morphology radio
sources 3C~433 and 4C~65.15: FR~IIs with asymmetric environments}

\author{B.P. Miller,\footnotemark[1] ~W.N. Brandt\footnotemark[1]}

\footnotetext[1]{Department of Astronomy and Astrophysics, The
Pennsylvania State University, 525 Davey Laboratory, University Park,
PA 16802, USA; {\it bmiller, niel@astro.psu.edu}}

\begin{abstract}

We present {\it Chandra} observations of the hybrid morphology radio
sources 3C~433 and 4C~65.15, two members of the rare class of objects
possessing an FR~I jet on one side of the core and an FR~II lobe on
the other. The \hbox{X-ray} spectrum of 3C~433 shows intrinsic
absorption (with a column density of \hbox{$N_{\rm H}\simeq
8\times10^{22}$ cm$^{-2}$}), such as is typical of FR~II narrow-line
radio galaxies. There is excess \hbox{X-ray} emission below 2~keV
containing contributions from diffuse soft \hbox{X-ray} emission
(likely hot gas with $kT\sim$1.2~keV) as well as from the nucleus. The
core of 3C~433 is extended in hard X-rays, presumably due to
\hbox{X-ray} emission from the inner-jet knot on the FR~I side that is
apparent in the radio map. It is possible that the \hbox{X-ray}
emission from this inner-jet knot is absorbed by the dust known to be
present in the host galaxy. The spectrum of 4C~65.15 can be modeled
with a simple power law with perhaps mild intrinsic absorption
(\hbox{$N_{\rm H}\simeq1.3\times10^{21}$ cm$^{-2}$}). \hbox{X-ray}
emission is detected at the bend in the FR~I jet. This \hbox{X-ray}
jet emission lies above the extrapolation from the high-frequency
radio synchrotron emission and has a spectral slope flatter than
${\alpha}_{\rm rx}$, indicating that the jet spectral energy
distribution is concave as with other FR~II quasar jets. Both 3C~433
and 4C~65.15 have unabsorbed X-ray luminosities, radio luminosities,
and optical spectra typically seen in comparable sources with FR~II
morphologies. Presumably the FR~I structure seen on one side in these
hybrid sources is generated by a powerful jet interacting with a
relatively dense environment.

\end{abstract}


\section{Introduction}

Radio-loud active galactic nuclei with extended radio emission may be
broadly classified as either edge-darkened or edge-brightened. As
defined following Fanaroff \& Riley (1974), FR~I sources show
initially prominent jets tapering off into dim and diffuse plumes,
whereas FR~II sources show tightly collimated jets terminating in
luminous hotspots and complex lobes. FR~I sources are typically less
radio luminous than are FR~IIs (Fanaroff \& Riley 1974), with the
dividing luminosity being an increasing function of host-galaxy
optical luminosity (Owen \& Ledlow 1994). At low redshift FR~Is tend
to inhabit richer groups than do FR~IIs (e.g., Zirbel 1997) although
this trend seems to vanish at moderate redshift ($z~\simgt$~0.3; e.g.,
Auger et al.~2008). Less than 1\% of radio sources possess both an
FR~I jet and an FR~II lobe on the opposite side of the core
(Gawro{\'n}ski et al.~2006). Gopal-Krishna \& Wiita~(2000) argue that
the structure of such ``HYMORS'' (HYbrid MOrphology Radio Sources) is
most plausibly due to the propagation of twin jets into an asymmetric
medium. Despite their rarity, such mixed sources are of significant
general interest to understanding the degree to which the surrounding
environment influences morphology and FR type.

X-ray observations provide a productive approach toward understanding
the nature of hybrid morphology sources, since the \hbox{X-ray}
properties of the nucleus and jets in FR~IIs are broadly distinct from
those of FR~Is. FR~Is show correlations between radio, optical, and
X-ray nuclear luminosity (extending to beamed sources), suggesting
that the majority of the core emission in FR~Is, from radio to X-ray
wavelengths, is produced in an unresolved synchrotron-emitting jet
(e.g., Chiaberge et al.~2000; Evans et al.~2006). The \hbox{X-ray}
spectra of FR~I sources typically do not show significant intrinsic
absorption, and it is unlikely that apparently unabsorbed FR~Is harbor
a luminous accretion disk hidden by heavy absorption, as they are not
particularly bright at IR wavelengths (e.g., M{\"u}ller et al.~2004);
this implies that most FR~Is are likely inefficient accretors (e.g.,
Chiaberge et al.~2000; Evans et al.~2006). FR~IIs display a wider
range of observed properties, presumably related to their inclination
to the line of sight. Some unification models (e.g., Urry \& Padovani
1995) for radio-loud FR~II sources connect narrow-line radio galaxies
(NLRGs), broad-line radio galaxies (BLRGs), and radio-loud quasars
(RLQs) based on viewing angle in a manner similar to radio-quiet
unification schemes: NLRGs are thought to be shrouded by a dusty
torus, BLRGs allow a clear view of the central engine, and RLQs are
seen closer to the axis of the jet. Indeed, BLRGs show optical
emission well in excess of that predicted by the FR~I radio-optical
correlation, presumably from the accretion disk (e.g., Varano et
al.~2004). \hbox{X-ray} studies also offer broad support for
unification models, as NLRGs generally show significant ($N_{\rm H} >
5\times10^{22}$ cm$^{-2}$) intrinsic absorption (e.g., Evans et
al.~2006) but RLQs typically do not (e.g., Belsole et
al.~2006)\footnote{Some high-redshift RLQs display significant
intrinsic \hbox{X-ray} absorption (e.g., Yuan et al.~2006 and
references therein).}; the \hbox{X-ray} spectrum of RLQs flattens as
inclination to the line of sight decreases, consistent with increasing
dominance from a beamed jet (e.g., Belsole et al.~2006).

X-ray observations also show clear differences between the properties
of the kpc-scale jets in FR~Is and FR~IIs. FR~I jets have flat or
convex spectral energy distributions (SEDs), with ${\alpha}_{\rm
ro}/{\alpha}_{\rm ox}\simlt 1$, and steep \hbox{X-ray} power-law
spectra ($\Gamma \simgt$ 2), consistent with a common synchrotron
origin for the radio-to-X-ray jet emission (e.g., Worrall \&
Birkinshaw 2006). However, the concave SEDs and flatter \hbox{X-ray}
spectra seen in FR~II quasar jets require a distinct origin for the
\hbox{X-ray} jet emission. One commonly discussed model is Compton
upscattering of cosmic microwave background photons (IC/CMB; e.g.,
Tavecchio et al.~2000), a particularly efficient method of
\hbox{X-ray} production at small inclination angles and large
redshifts; however, the predicted dominance of \hbox{X-ray} jets over
core emission at high redshifts (Schwartz 2002) has not yet been
observed (e.g., Bassett et al.~2004; Lopez et al.~2006), and there are
additional complications to the IC/CMB model (e.g., Hardcastle
2006). An appealing alternative is that the X-ray jet emission in RLQs
with concave SEDs may arise from a second population of highly
energetic synchrotron-emitting electrons (e.g., Atoyan \&
Dermer~2004).

We report here on recent {\it Chandra} observations of the hybrid
radio sources 3C~433 and 4C~65.15. 3C~433 is a low-redshift
($z$~=~0.102) NLRG with highly unusual extended radio emission (e.g.,
van Breugel et al.~1983), giving it an ambiguous FR~I/II morphological
classification (Wills et al.~2002). There is a luminous southern FR~II
lobe that includes a primary hotspot and bright outer ridges and a
knotty northern FR~I jet ending in a diffuse plume that stretches
perpendicularly to the east.\footnote{For simplicity we will refer
throughout to the ``FR~II lobe'' and ``FR~I jet'' in each object;
these classifications should be understood to refer to the dominant
type of structure on a given side.} The complex southern lobe contains
considerable cold gas, as indicated by HI measurements (Mirabel et
al.~1989; Morganti et al.~2003). Black et al.~(1992) resolve the core
into two components (C2 and C1) separated by only $\sim$1 kpc; the
northern core component (C1) is most likely an inner jet
knot. 4C~65.15 ($z$~=~1.625) was noted as a possible hybrid object by
Gopal-Krishna \& Wiita (2000); the radio image shows a southern jet
lacking a primary hotspot (Lonsdale et al.~1993) that peaks in
brightness as it abruptly bends at a right angle and then gradually
expands and fades away to the west, somewhat similar to the striking
structure of the FR~I jet of 3C~433. The (strongly polarized) northern
FR~II hotspot is much brighter than the nucleus at 8.5~GHz. The Sloan
Digital Sky Survey (SDSS; York et al.~2000) spectrum of 4C~65.15 shows
typical quasar C~IV, C~III, and Mg~II broad emission lines; 4C~65.15
also has associated narrow absorption blueward of the C~IV emission
line (Vestergaard 2003). Both objects have radio luminosities
consistent with those of comparable FR~II sources. Both also are
lobe-dominated; while the RLQ 4C~65.15 is presumably inclined closer
to the line of sight than the NLRG 3C~433, beaming or projection
effects alone cannot explain the hybrid structure in either case. The
primary scientific goal for this project is to determine the
fundamental nature of these objects through categorizing their nuclear
and jet \hbox{X-ray} properties as either FR~I, FR~II, or mixed.

This paper is organized as follows: $\S$2 presents the X-ray
observations and relevant images, $\S$3 describes analysis of the
3C~433 data, $\S$4 describes analysis of the 4C~65.15 data, $\S$5
discusses the results in the context of known properties of FR~Is and
FR~IIs, and $\S$6 summarizes the main conclusions of this work. A
standard cosmology with \hbox{$H_{0}$ = 71 km s$^{-1}$~Mpc$^{-1}$},
${\Omega}_{M}$~=~0.27, and ${\Omega}_{\Lambda}$~=~0.73 is assumed
throughout. This choice results in luminosity distances of 465 and
12160 Mpc and angular-distance scales of 1.86 and 8.56 kpc
arcsec$^{-1}$ for 3C~433 and 4C~65.15, respectively. The Galactic
column density toward 3C~433 (${\alpha}_{2000}$~=~21~23~44.5,
${\delta}_{2000}$~=~+25~04~27) is 1.19$\times$10$^{21}$ cm$^{-2}$;
toward 4C~65.15 (${\alpha}_{2000}$~=~13~25~29.7,
${\delta}_{2000}$~=~+65~15~13) it is 1.95$\times$10$^{20}$
cm$^{-2}$. Unless otherwise noted, errors are given as 90\% confidence
intervals for one parameter of interest (${\Delta}{\chi}^{2}$~=~2.71).

\section{Observations}

3C~433 was observed by {\it Chandra} on 2007 Aug 28 using ACIS-S3 in a
standard 1/2 subarray. After Good Time Interval (GTI) filtering the
exposure time was 37.2 ks, and the count rate from the core was 0.076
counts s$^{-1}$; there are $\sim$2800 total counts in the core. The
source light curve does not show significant variability. Radio and
optical data help place the {\it Chandra} results in context. We make
use of high-resolution radio maps of 3C~433 created from archival {\it
VLA} data at two frequencies: a 1.5~GHz image with a resolution of
1.0$''$ (from observations conducted on 1986 Apr 25) illustrates the
large-scale radio structure, while a 8.3~GHz image with a resolution
of $0.25''$ based on data presented in Black et al.~(1992) reveals the
structure of the inner jet. An {\it HST} WFPC2 image was retrieved
from the MAST archives\footnote{Multimission Archive at STScI:
http://archive.stsci.edu/index.html}; these data were previously
discussed by de Koff et al.~(1996), who noted the large-scale dust
absorption features in the host galaxy. An adaptively smoothed 0.5--2
keV image is shown in Figure 1a, overlaid with 1.5 GHz radio
contours. The smoothed image was generated using the CIAO task {\it
csmooth} with a minimum significance (S/N ratio) of 2 and a maximum
significance of 5. There is extended \hbox{X-ray} emission within the
southern lobe and to the north on either side of the radio jet; these
structures are also apparent in unsmoothed images. Closer
investigation reveals that the \hbox{X-ray} nucleus is resolved by
{\it Chandra} (see $\S$3).

4C 65.15 was observed by {\it Chandra} on 2007 Jul 20 using ACIS-S3 in
full-frame mode. After GTI filtering the effective exposure time was
35.8 ks and the count rate in the core was 0.044 counts s$^{-1}$;
there are $\sim$1600 total counts in the core. The source light curve
shows mild ($\sim$20\%) variability on ks timescales. Analysis of the
jet and environment of 4C~65.15 was aided by 8.5 GHz radio data with a
resolution of $0.25''$ (from observations conducted on 1999 Aug 9),
4.9~GHz data with a beamsize of 0.47$''\times0.31''$ at position angle
13$^{\circ}$ (from observations conducted on 1983 Sep 19), and optical
data in the form of an $r$-band SDSS image. An adaptively smoothed
0.5--8 keV image is shown in Figure 2, overlaid with radio
contours. The smoothed image was created from an image rebinned to
0.25$''$ pixels from which the pipeline pixel randomization had been
removed, and was generated using the CIAO task {\it csmooth} with a
minimum significance of 1.5, a maximum significance of 5, and an
initial minimum smoothing scale of 1.3 pixels. The \hbox{X-ray} core
is unresolved. \hbox{X-ray} jet emission is detected at the location
of the bend on the FR~I side of the nucleus and appears to extend
along the jet. There appears to be \hbox{X-ray} emission near the
tail\footnote{Lower frequency maps (e.g., Reid et al.~1995) show
diffuse emission to the northwest of the nucleus, suggesting it is
possible that the jet continues to curve and fan out.} of the FR~I jet
and near the FR~II lobe, $\sim$5 counts in both cases; the local
background in comparable-sized regions a similar distance from the
core is $\sim$1.5 counts, so these are only marginal detections.

X-ray spectra were extracted from the nuclear region and from other
areas of interest for both sources, and models were fit using XSPEC 12
(Arnaud 1996). After examining the spectra, we preferred to fit the
ungrouped spectrum for 3C~433 (using the $C$-statistic; Cash 1979) to
model the low-count region below 2 keV better. Such considerations do
not apply to 4C~65.15, so this spectrum was grouped to contain 15
counts per bin and fit using the $\chi^{2}$ statistic. All fits were
conducted over the 0.3--8 keV energy band, and all models include
fixed Galactic absorption.

\section{3C 433}

The nuclear \hbox{X-ray} spectrum of 3C~433 was extracted from a
circular region with a radius of 5 pixels ($\simeq$2.5$''$), centered
on the peak flux. The spectrum (Figure 3a) shows the intrinsic X-ray
absorption expected in a NLRG. A simple power-law model with intrinsic
neutral absorption has best-fit parameters for the column density of
$N_{\rm H} = 7.17\times10^{22}$ cm$^{-2}$ and for the photon index of
$\Gamma$ = 1.27 but does not provide a satisfactory fit, as indicated
by the large positive residuals below 2 keV and the high $C$-statistic
value (691 for 523 degrees of freedom; 99.97\% of Monte Carlo
simulations conducted using the XSPEC {\it goodness} command have
lower $C$-statistic values, indicating the fit is poor). The spectrum
can be satisfactorily fit with a partial-covering model with a
covering fraction of 0.993 ($N_{\rm H} = 8.64\times10^{22}$ cm$^{-2}$,
$\Gamma$~=~1.44, $C$-statistic/d.o.f. = 596/522, 50.10\% of
simulations have lower \hbox{$C$-statistic} values, indicating an
acceptable fit). Table 1 lists parameters and errors for this and the
following models. The excess soft emission can also be accommodated
through an additional emission component, either a power-law
(\hbox{${\Gamma}_{\rm unabs}~=~2.2\pm0.8$}) or thermal bremsstrahlung
($kT = 1.2^{+14.2}_{-0.6}$~keV; note the upper limit for the
temperature is poorly constrained); this unabsorbed component would
contain $\sim$80 counts, primarily in the soft band. The total 0.5--8
keV model flux, dominated by the absorbed component, is
$1.84\times10^{-12}$ erg cm$^{-2}$ s$^{-1}$. There is a 34~ks {\it
ASCA} SIS spectrum obtained on 1997 May 28 (PI Yamashita) with a
0.5--8~keV model flux of $2.75^{+1.92}_{-1.33}\times10^{-12}$ erg
cm$^{-2}$ s$^{-1}$, consistent with this {\it Chandra} observation.

There is diffuse soft \hbox{X-ray} emission surrounding the nucleus
and extending particularly toward the northeast and northwest (see
Figure 1c) out to $\sim$5--6$''$ ($\sim$9--11 projected kpc). It
appears most likely that this soft \hbox{X-ray} flux is associated
with hot gas such as is often observed around FR~II radio sources. The
gas is not distributed in a symmetric halo, but non-spherical
distributions have been found in {\it Chandra} observations of other
radio galaxies (e.g., Kraft et al.~2005). The {\it HST} image (Figure
1b) shows a faint point source northwest of the host galaxy; were this
the source of the X-rays to the northwest, the X-ray/optical flux
ratio would be consistent with the object being a background AGN
(Maccacaro et al.~1988), although an optical spectrum is necessary for
conclusive classification. Despite the sparse counts the soft
\hbox{X-ray} emission in this northwestern area appears extended,
leading us to favor hot gas as the emission source. The best-fit
temperature for a thermal bremsstrahlung model applied to the
northwest diffuse \hbox{X-ray} emission is $kT=1.37^{+1.66}_{-0.59}$
keV. The volume corresponding to the X-ray emission is difficult to
measure accurately but may be approximated as a sphere with a radius
of 2.5$''$ (4.7 kpc). Neglecting line emission (which contributes
significantly to the soft \hbox{X-ray} emission below 1~keV), a gas
cloud with approximately solar abundances would be required to have a
density of $n\sim$~0.06~cm$^{-3}$ to account for the observed
\hbox{X-ray} flux, corresponding to a total mass of
6.6$\times$10$^{8}$~M$_{\odot}$ and an ideal-gas pressure of
1.5$\times$10$^{-10}$ dynes cm$^{-2}$. The northeastern diffuse X-ray
emission appears to occupy a slightly smaller volume, and could have
similar density and pressure with about half the total mass as the gas
to the northwest. The hot gas would be overpressured with respect to
typical IGM temperatures and densities, suggesting either that we are
observing it at a favorable time before it disperses (see also
$\S$5.2) or else there is a quasi-continuous source of heating,
perhaps related to the northern jet.

We can utilize spatial analysis of the soft \hbox{X-ray} emission in
the vicinity of the core to determine whether the low-energy X-ray
spectrum originates primarily in the core or is dominated by the
diffuse emission. A 0.3--1~keV image was constructed with the pipeline
pixel randomization removed and with 0.1$''$ pixel binning. Figure 3b
shows a radial profile extracted from this image using circular annuli
compared with a 1~keV PSF calculated with MARX.\footnote{MARX is a
{\it Chandra} ray-trace simulator; see http://space.mit.edu/ASC/MARX/}
The 0.3--1~keV emission within the spectral-extraction region contains
contributions from both nuclear and diffuse emission; from the PSF
normalization, $\sim$50\% of the counts to $\simeq$~2.5$''$ are from
the core and $\sim$50\% from the diffuse emission. For the double
power-law model, $\sim$80\% of the 0.3--1~keV counts are from the
unabsorbed component (with similar results when the low-energy
emission is modeled as thermal bremsstrahlung), with the absorbed
component only contributing $\sim$20\% of the counts over this energy
range. This suggests $\sim$30\% of the 0.3--1~keV counts originate in
the core and are not associated with the absorbed spectral component;
this may be soft \hbox{X-ray} emission from an unresolved small-scale
jet. Separate consideration of the soft \hbox{X-ray} emission to the
north and south of the nucleus indicates the soft emission does not
skew strongly to the north, implying the C1 inner jet knot contributes
only a small fraction of the soft emission within the spectral
extraction region; however, the uncertainties with so few counts are
large.

There are sufficient counts over the entire {\it Chandra} spectrum to
resolve the core region on sub-arcsecond scales and to determine
directly whether the inner jet knot C1 is detected in X-rays. We
performed a maximum-likelihood reconstruction (cf. $\S$2 of Townsley
et al.~2006) on images from which the pipeline pixel randominzation
had been removed and that were binned to 0.1$''$ pixels. These spanned
several different energy ranges, using appropriate PSFs calculated
with MARX. The \hbox{X-ray} core is indeed extended toward the north,
with emission in excess of that expected from a point source observed
at approximately the same position angle and distance as C1 is
relative to C2 (see Figures 4a and 4b). The C1-linked \hbox{X-ray}
emission is most apparent in hard-band images and is therefore not
directly related to the diffuse soft \hbox{X-ray} emission. The
observed extension in the X-ray core is not an artifact of the mirror
pair 6 misalignment (cf. $\S$4.2.3 of The {\it Chandra} Proposers'
Observatory Guide), as it persists at energies below $\sim$4~keV. It
is unlikely that the core, observed on-axis and consisting of
$\sim$2800 counts, would be significantly distorted in one direction
due solely to statistical noise; even were such to occur, it is
further unlikely that the angle and position would randomly align so
closely with C1. The 2--6~keV radial profile also shows a
statistically significant excess of X-ray emission beyond that
expected from a point source to the north of the X-ray core (Figure
4c). Adding an additional point source 0.6$''$ from the core provides
a much improved (but not exact) match to the overall profile. The
normalization is $\simeq$10\% that of the core, suggesting that
C1-linked \hbox{X-ray} emission contributes a few hundred counts to
the overall \hbox{X-ray} spectrum. There is no indication that the
hard \hbox{X-ray} component in the overall spectrum itself requires
multiple power-laws for an acceptable fit, so the C1-linked
\hbox{X-ray} spectrum appears to be broadly similar to that of the
core.

No Fe K$\alpha$ emission line is detected in the \hbox{X-ray} spectrum
of 3C~433. The fit is not significantly improved by adding a Gaussian
with fixed rest-frame energy 6.4 keV and fixed width 0.1 keV; the
rest-frame upper limit to the equivalent width of any Fe K$\alpha$
emission is 85 eV. Permitting the line energy to vary does not suggest
any emission from ionized iron. Inspection of the spectrum reveals a
marginal feature at rest-frame 8.2 keV (equivalent width $\simeq$ 150
eV), but the model fit is not significantly improved after adding this
component and in any event there is no obvious physical origin for
such emission here. The lack of iron emission is somewhat atypical for
X-ray spectra of heavily absorbed (FR~II) radio galaxies, which often
show Fe K$\alpha$ line emission of $\sim100-300$~eV equivalent width
(e.g., Evans et al.~2006); perhaps the somewhat lower intrinsic
absorption in 3C~433 results in emission from a narrow Fe K$\alpha$
line being diluted below detectability.

The southern lobe of 3C~433 contains diffuse \hbox{X-ray} emission
that is easily seen in the soft-band unsmoothed image and appears as a
curving band along the east side of the southern lobe in the 0.5--2
keV smoothed image. There are $\sim120\pm$14 counts (1$\sigma$ errors)
above background in the 0.3--8 keV band, with a 0.5--8 keV model flux
of $2.3\times10^{-14}$~erg~cm$^{-2}$~s$^{-1}$. Similarly sized regions
extracted from east, south, and west of the lobe do not contain a
statistically significant excess of counts above background. The
extent of lobe emission above 2~keV is unclear; smoothed hard-band
images suggest extended emission preferentially located southwest of
the nucleus, but this cannot be confirmed in unsmoothed images.

\section{4C~65.15}

There are no bright inner radio-jet knots in 4C~65.15 and no
indication that the \hbox{X-ray} nucleus of 4C~65.15 is extended. The
nuclear \hbox{X-ray} spectrum was extracted from a circular region
with a radius of 2.4$''$; the \hbox{X-ray} jet emission is located
outside this area. The initial model for 4C~65.15 consisted of an
unabsorbed power-law. The best-fit photon index is $\Gamma =
1.89\pm0.07$ and the fit is acceptable, with $\chi^{2}$ = 55.75 for 82
degrees of freedom. The 0.5--8~keV model flux is $2.55\times10^{-13}$
erg cm$^{-2}$ s$^{-1}$. There is a 6~ks {\it ROSAT} PSPC observation
obtained on 1992 Nov 30 (PI Laor) with a count rate implying an
extrapolated 0.5--8~keV flux of
$2.38\pm0.36\times10^{-13}$~erg~cm$^{-2}$~s$^{-1}$, consistent with
this {\it Chandra} observation. The model was slightly improved by
adding intrinsic absorption with a best-fit neutral column density of
$N_{\rm H} = 1.31^{+1.43}_{-1.25}\times10^{21}$~cm$^{-2}$, with the
photon index adjusting to $\Gamma = 1.97^{+0.11}_{-0.05}$. Note that
the 90\% confidence region for the column density is barely above
zero. The $\chi^{2}$/dof for this model is 52.75/81, a decrease with
an $F$-test probability of occuring by chance of 3.5\%. This fit is
shown in Figure~5. Some \hbox{X-ray} absorption is perhaps plausible
in light of the associated narrow C~IV absorption in 4C~65.15 and the
general tendency for \hbox{X-ray} and UV absorption to be linked
(e.g., Brandt et al.~2000; Gallagher et al.~2001).

Adding an Fe K$\alpha$ emission line with a fixed rest-frame energy of
6.4 keV and a fixed width of 0.1 keV did not improve the fit, and the
spectrum shows no noticeable excess emission at that energy (the
rest-frame upper limit to the equivalent width is 101 eV). Permitting
the energy of the line to vary gives a best-fit value of rest-frame
7.10~keV; fitting to the ungrouped spectrum (using the $C$-statistic)
also gives an energy of 7.06~keV for a 34 eV (rest-frame) equivalent
width line, but the 90\% confidence interval for the line
normalization includes zero. We conclude there is no significant iron
emission detected in this spectrum.

The \hbox{X-ray} emission in the FR~I jet in 4C~65.15 contains
sufficient photons for basic spectral modeling (31 counts in the
0.3--8 keV band) using the $C$-statistic and permitting only one
spectral-shape free parameter. A power-law model (plus fixed Galactic
absorption) gives $\Gamma = 1.17^{+0.41}_{-0.49}$ and a 0.5--8 keV
model flux of $7.9\times10^{-15}$ erg cm$^{-2}$ s$^{-1}$. A thermal
model is not a very good fit for reasonable temperatures (the best-fit
is $kT\sim$200~keV) and is physically unlikely for jet-related X-ray
emission.

There is apparent excess \hbox{X-ray} emission to the northeast of the
core of 4C~65.15 located near the northern FR~II radio lobe, but the
paucity of counts above background ($\sim$3.5 net counts over 0.5--8
keV) makes this only a marginal detection. \hbox{X-ray} emission from
hotspots in FR~II sources can often be successfully explained with
synchrotron self-Compton models, although low-luminosity cases may be
simple synchrotron. We can estimate the expected \hbox{X-ray} emission
from the FR~II lobe of 4C~65.15 using a radio-to-X-ray flux ratio
typical for other FR~II hotspots (cf. Table 3 from Hardcastle et
al.~2004) and obtain a predicted 0.5--8~keV \hbox{X-ray} count rate of
$\sim(4-80)\times10^{-5}$~counts~s$^{-1}$, or $\sim1.5-30$~counts in
35.8~ks. The lower end of this range is more representative of
luminous hotspots, such as that of 4C~65.15, and is broadly consistent
with the observed counts.

There is also apparent excess \hbox{X-ray} emission within the tail of
the southwest jet (again $\sim$3.5 net counts over 0.5--8 keV, a
marginal detection). It is not immediately clear whether such
emission, if genuine, is associated with a terminal hotspot or an
outer jet knot. In the context of a hotspot interpretation, the low
radio flux of this feature suggests an expected 0.5--8~keV
\hbox{X-ray} count rate of $\sim(3-50)\times10^{-6}$~counts~s$^{-1}$,
or $\sim0.1-1.8$~counts in 35.8~ks (calculated as above). Here the
higher end of this range is more representative for hotspots of lower
luminosities, but even so this would be a somewhat \hbox{X-ray} bright
hotspot. If instead the X-ray emission arises in a jet feature, the
ratio of X-ray-to-radio flux is perhaps somewhat less than in the
bend, plausibly decreasing along the jet such as is often
observed. Deeper \hbox{X-ray} observations would be required to
conduct more quantitative analysis.

\section{Discussion}

\subsection{Jet emission}

The inner jet in 3C~433 (C1) is detected in \hbox{X-rays}, but it is
difficult to be quantitative about its characteristics due to its
close proximity to the nucleus. The jet/core flux ratio for C1 of
$\sim$10\% is somewhat larger than for typical FR~II quasar jet knots,
which generally have values of 1--8\% (e.g., Marshall et
al.~2005). The radio-to-X-ray spectral slope of ${\alpha}_{\rm
rx}\sim$0.8 is also flatter than the ${\alpha}_{\rm rx}\sim$0.9--1.0
that is typical for FR~II quasar knots (e.g., Marshall et
al.~2005). The inner knot C1 appears to be relatively \hbox{X-ray}
luminous (note the comparisons do not take into account the distance
of knots from the core). At the large inclinations indicated for this
NLRG, the IC/CMB process should not contribute significantly to jet
\hbox{X-ray} emission, and instead the \hbox{X-ray} jet emission may
be synchrotron dominated. The data do not permit differentiation
between simple synchrotron and multiple component models.

The \hbox{X-ray} spectrum of the inner jet appears to be dominated by
hard-band emission, perhaps suggesting that C1 is absorbed ($N_{\rm
H}>1.7\times10^{22}$~cm$^{-2}$) even at $\sim1$~kpc distance from
the core. The ``torus'' of absorbing material presumably responsible
for obscuring the core typically is believed not to extend to such
distances (e.g., Maiolino \& Risaliti 2007). However, this location is
well within the scale of the dust structures seen in the host galaxy,
so C1 may happen to lie along a line of sight that passes through a
dense dust cloud or lane; Figure 1b provides tentative support for
this hypothesis. de Koff et al.~(2000) estimate the mass of dust in
3C~433 as 10$^{5.7}$~M$_{\odot}$ based on absorption maps and
10$^{8}$~M$_{\odot}$ based on emission measured by IRAS, so there may
be sufficient material to account for the inferred absorption if a
clump covers the \hbox{X-ray} emitting region of the inner jet.

The nuclear soft \hbox{X-ray} emission might itself contain a
contribution from an unresolved small-scale jet, as discussed in
$\S$3. Some support for this idea is provided by the agreement of the
core radio/X-ray luminosity ratio with those of unabsorbed
``jet-related'' components in the FR~IIs studied by Evans et
al.~(2006). The 3C~433 X-ray spectrum suggests that such nuclear
jet-related emission is not strongly absorbed, contrasting with the C1
jet knot; indeed, strong absorption of the nuclear jet-related
emission would render detection as a distinct spectral component
difficult. Further, if the unabsorbed luminosity for this nuclear
jet-related component were calculated from the observed nuclear
soft-band flux but assuming the C1 column density, then the X-ray
luminosity would exceed the expected FR~II trend based on the radio
core luminosity (although there are substantial uncertainties in the
measurements as well as scatter in the correlation). Since the dust in
the host galaxy of 3C~433 appears to be distributed in a patchy
manner, the column densities associated with this dust may well vary
on scales $\simlt1$~kpc.

The jet in 4C~65.15 has a clear detection in X-rays and radio flux
measurements at several frequencies. The \hbox{X-ray} photon index of
$\Gamma = 1.17^{+0.41}_{-0.49}$ from the power-law model is consistent
with the $\Gamma~\sim$~1.1--1.7 found by Sambruna et al.~(2004) for
the brightest \hbox{X-ray} knots in their {\it Chandra} and {\it HST}
survey of core-dominated FR~II quasars with known radio jets. It does
not match the \hbox{X-ray} spectra of prominent knots in FR~I jets,
which are generally significantly steeper with $\Gamma~\simgt$~2
(e.g., 3C~66B: Hardcastle et al.~2001; 3C~31: Hardcastle et al.~2002;
M~87: Marshall et al.~2002; Cen~A: Hardcastle et al.~2003; B2~0755+37:
Parma et al.~2003). The spectral energy distribution of the
\hbox{X-ray} emitting jet region is plotted in Figure 6. The radio
points are summed over the resolved features at the jet bend. The
radio-to-X-ray spectral slope is ${\alpha}_{\rm rx}$ = 0.96;
unfortunately the SDSS upper limit does not impose useful constraints
upon ${\alpha}_{\rm ro}$ or ${\alpha}_{\rm ox}$. The high-frequency
radio spectral slope is steeper; between 5 and 15 GHz, ${\alpha}_{\rm
r}$ = 1.08. Consequently, the high-frequency radio spectral slope
predicts synchrotron \hbox{X-ray} emission (for simple models,
assuming the population of electrons extends to sufficiently high
energies) with a flux lower than the \hbox{X-ray} jet emission that is
actually detected. Moreover, the measured \hbox{X-ray} spectral index
is much flatter than ${\alpha}_{\rm rx}$; the 90\% confidence upper
limit from spectral fitting for ${\alpha}_{\rm x}$ is 0.6. It appears
most likely that the jet X-ray emission is not an extension of the
synchrotron component responsible for the radio emission, although
deeper optical observations would be helpful for better understanding
the shape of the jet SED.

The \hbox{X-ray} and radio surface-brightness profiles within the
4C~65.15 jet bend do not appear to align; Figure 2 suggests the
\hbox{X-ray} emission peaks slightly upstream of the maximum radio
brightness. (The paucity of counts as well as the limited angular size
of the region of interest in the jet bend restrict the usefulness of
more detailed spatial analysis.) Such an offset would make it less
likely that the IC/CMB process dominates the jet \hbox{X-ray}
emission, since the low-energy electrons involved should not
congregate upstream of the peak radio synchrotron emission. This
conclusion might be reached independently through consideration of the
lobe-dominated nature of 4C~65.15, which constrains the inclination
and hence limits the efficiency of IC/CMB emission. A model involving
a second high-energy population of \hbox{X-ray} synchrotron emitting
electrons could accommodate both any offset and a concave jet SED. The
high radio luminosity, broad-line quasar classification, and
apparently concave jet SED of 4C~65.15 are similar to the properties
of typical FR~II quasars, suggesting that the distorted southern radio
structure in 4C~65.15 reflects an external influence, most likely a
dense surrounding environment disrupting the outer jet.

\subsection{Environment}

3C~433 appears to reside in a group environment. Zirbel (1997) found a
background-corrected group richness of 12.6$\pm$4.4 for 3C~433; the
mean richness among low redshift FR~IIs in that study was 5.8 (scatter
5.6), and 76\% had a richness of $<$10. There is a close companion
$\sim14$ projected kpc to the northeast; although the redshift of this
nearby galaxy is unknown, its angular size is consistent with being at
the same distance as 3C~433. There is another galaxy $\sim17$
projected kpc to the north and an additional galaxy $\sim34$ projected
kpc to the southwest, as well as an optical point source $\sim4''$
northwest of the optical nucleus of 3C~433 whose nature is
unclear. There is evidence that 3C~433 is either experiencing tidal
forces driving star formation or else has undergone a recent minor
merger: Wills et al.~(2002) identify a young stellar population in
3C~433 from UV/optical spectral analysis, and the dust structure
suggests 3C~433 has been disturbed (de Koff et al.~2000) in some
manner. Such activity could plausibly generate asymmetries in the
surrounding intergalactic medium.

The SDSS image of 4C~65.15 shows a handful of nearby optical sources,
but they are too faint to have SDSS spectral coverage and so their
redshifts are uncertain. There is no overdensity of nearby sources
(with $m_{\rm r} <$ 23) in the vicinity of 4C~65.15 on scales of
90$''$, 60$''$, or 30$''$, nor are there more galaxies (resolved
sources: type=3) near 4C~65.15 on these same scales.

Both 3C~433 and 4C~65.15 display bends in the jet on the FR~I side of
the nucleus. While bends in radio jets are not uncommon, the change in
direction is particularly abrupt ($\sim90^{\circ}$) for both 3C~433
and 4C~65.15. 3C~433 likely lies with its jet axis nearly in the plane
of the sky, as indicated by its large lobe-to-core flux ratio and NLRG
status, and so the observed bend should closely correspond to the
physical change in direction. The bend in 4C~65.15 may be exaggerated
due to an orientation closer to the line of sight. Judging from the
core-to-lobe flux ratio and from the core radio-to-optical luminosity
ratio (e.g., Wills \& Brotherton 1995), 4C~65.15 is inclined at
$\sim35^{\circ}\pm10^{\circ}$ to the line of sight; this is the lower
limit for the deprojected bending angle (e.g., see $\S$3 of Jorstad \&
Marscher~2004). The jet structure in 3C~433 and 4C~65.15 is suggestive
of an environmental interaction redirecting the jet, consistent with a
dense clumpy medium on the FR~I side such as might also produce the
hybrid structure through decelerating the jet closer to the core. An
interesting possibility for producing X-ray emission at the site of an
abrupt jet bend (such as is observed in 4C~65.15) is suggested by
Worrall \& Birkinshaw (2005) for 3C~346: interaction between the jet
and a wake created by a companion galaxy moving through the
intergalactic medium generates an oblique shock, redirecting the jet
and powering X-ray synchrotron emission. It is perhaps noteworthy that
3C~433 has been included in lists of ``X-shaped'' radio sources; while
there have been suggestions that such morphologies arise from the
merger of supermassive black holes (e.g., Merritt \& Ekers 2002), it
has been convincingly argued (e.g., Worrall et al.~1995; Kraft et
al.~2005; Cheung 2007) that hydrodynamic backflows from the jet within
an asymmetric medium can naturally produce ``X-shaped'' structure.

\subsection{Comparison to other radio sources}

There is a correlation between core radio and \hbox{X-ray} luminosity
for FR~I sources, presumably reflecting jet dominance of the nuclear
emission from radio through \hbox{X-ray} frequencies. FR~II sources
tend to have higher (unabsorbed) \hbox{X-ray} luminosities at a given
core radio luminosity, perhaps because their \hbox{X-ray} emission
includes a contribution from Compton upscattering of disk photons in a
hot corona (e.g., Evans et al.~2006). Both 3C~433 and 4C~65.15 have
nuclear (unabsorbed) \hbox{X-ray} luminosities that lie above the FR~I
correlation, consistent with other FR~II sources (Figure 7). The
nuclear X-ray spectra are also FR~II in nature (intrinsic absorption
for the NLRG 3C~433, a power law for the RLQ 4C~65.15), and the jet
SED of 4C~65.15 is consistent with those of other FR~II quasar
jets. We conclude that the hybrid sources 3C~433 and 4C~65.15 should
be regarded as FR~II objects in which a particularly dense environment
has induced FR~I-like jet structure.

There are only a handful of known hybrid morphology sources possessing
high-resolution \hbox{X-ray} data. Gopal-Krishna \& Wiita (2000) list
six HYMORS, from which three (PKS~0521$-$365, PG~1004+130, and
S5~2007+777) have {\it Chandra} coverage. Birkinshaw et al.~(2002)
detected \hbox{X-ray} emission from the FR~I jet of the BL Lac
PKS~0521$-$365 which could be satisfactorily interpreted with a
synchrotron model, and also detected the FR~II hotspot in X-rays with
a flux similar to that expected from previous observations of hotspots
in FR~IIs. Miller et al.~(2006) found an \hbox{X-ray} counterpart
slightly upstream of the radio FR~I jet in the broad absorption line
RLQ PG~1004+130, with a flat photon index and a concave SED more
typical of FR~II quasar jets than FR~I jets. Sambruna et al.~(2008)
presented {\it Chandra} observations of the BL Lac S5~2007+777; the
FR~I jet is detected in X-rays and displays properties similar to
those of other FR~II quasar jets. Another possible hybrid source with
{\it Chandra} coverage is the BLRG 3C~17, which displays an
edge-brightened northern lobe and a curving jet to the southeast that
features a bright inner knot, an abrupt bend, and then an expanding
tail (Morganti et al.~1999). Two X-ray emitting features in the jet
were discovered by Massaro et al.~(2008). \hbox{X-ray} emission in the
inner knot is consistent with a single-component synchrotron spectrum,
although an IC/CMB interpretation is also possible. The outer
\hbox{X-ray} knot, located at the bend, appears to have a hard
\hbox{X-ray} spectrum, and a synchrotron model following the suggested
UV cutoff predicts lower \hbox{X-ray} flux than is observed,
indicating the SED is likely concave. Both PG~1004+130 and 3C~17 would
be classified as FR~II sources based on their radio luminosities and
optical broad line emission, and their jet SEDs match more closely to
FR~II quasar jets than to typical FR~I sources. The radio/optical
properties of the BL Lac objects PKS 0521$-$365 and S5 2007+777 are
intermediate between FR~I and FR~II, and for these sources the
prominence of the FR~I jet may reflect their low inclination to the
line of sight.

\section{Conclusions}

{\it Chandra} observations of the hybrid morphology radio sources
3C~433 and 4C~65.15 reveal that they have \hbox{X-ray} properties
consistent with those of comparable FR~II sources, supporting the
hypothesis that the apparent FR~I structure in the jet arises from
environmental interactions. In particular, we find that:

1. The narrow-line radio galaxy 3C~433 displays nuclear X-ray
absorption with a column density of $N_{\rm H} = 8.3\times10^{22}$
cm$^{-2}$, similar to other NLRGs and in agreement with predictions
for high-inclination FR~II sources from some unification models. The
broad-line radio-loud quasar 4C~65.15 shows at most mild intrinsic
absorption, similar to other RLQs and in agreement with predictions
for intermediate-inclination FR~II sources from some unification
models.

2. The unabsorbed nuclear \hbox{X-ray} luminosities for both 3C~433
and 4C~65.15 lie along the track populated by FR~II sources when
plotted against core radio luminosity and do not fall along the
radio/X-ray luminosity correlation followed by FR~Is (Evans et
al.~2006).

3. The knot/core and \hbox{X-ray}/radio luminosity ratio of the inner
jet in 3C~433 (the C1 component) are rather high, but the close
proximity to the core makes it difficult to assess the full SED. The
\hbox{X-ray} jet in the RLQ 4C~65.15 has a flat photon index ($\Gamma
\sim 1.2$) that is flatter than the radio-to-X-ray spectral slope,
suggesting the SED is concave. Flat \hbox{X-ray} spectral slopes and
concave SEDs are characteristic of FR~II quasar jets and distinguish
them from FR~I jets.

{\it Chandra} observations of additional hybrid morphology objects
would be useful in clarifying whether most such sources can be
characterized as intrinsically FR~IIs with a one-sided FR~I jet
structure generated through environmental interactions. If so, it
might be expected that FR~IIs residing within a particularly dense but
symmetric environment could display FR~I-like jet structure on both
sides of the core. Perhaps the few known broad-line FR~Is such as
2114+820 (Lara et al.~1999) are such objects; 2114+820 has
radio-core/X-ray luminosity properties consistent with FR~II sources
(Figure 7). Deep radio observations sensitive to extended emission are
necessary to detect FR~I jets beyond the local universe; Heywood et
al.~(2007) carried out a VLA survey of 18 radio-luminous broad-line
quasars with extended structure and moderate redshifts
($z$~=~0.36--2.5) and discovered that 4--6 (22--33\%) showed apparent
FR~I morphologies. There should also be varying degrees of
hybridization if this interpretation is correct. 3C~433 itself could
be considered somewhat of an intermediate case between a source with
two-sided FR~I structure and one with an FR~I jet opposing an FR~II
lobe, since its extended and complex southern lobe is suggestive of
jet disruption prior to the outer hotspot.

It would also be interesting to examine hybrid morphology radio-loud
quasars known to be inclined with the FR~I side pointed toward the
observer. Such objects should be intrinsically unabsorbed, so if the
density of the hypothesized ``frustrating'' medium matches that
suggested as necessary for jet confinement or disruption (e.g., De
Young 1993; Carvalho 1998) it might be directly detectable as
absorption against the bright nucleus. 

\acknowledgments

We thank the anonymous referee for constructive suggestions that
improved the paper. We thank Leisa Townsley and Pat Broos for
assistance with maximum-likelihood image reconstruction, Doug Gobeille
for reducing and analyzing archival L-band VLA data of 3C~433, Ed
Fomalont and the NRAO VLA Archive Survey\footnote{The NVAS can
currently be browsed through http://www.aoc.nrao.edu/$\sim$vlbacald/}
team for reducing and analyzing C-band and X-band observations of 3C
433 and 4C 65.15, and Dan Harris and Gordon Garmire for productive
conversations. We acknowledge support for this work under {\it
Chandra} \hbox{X-ray} Center grant GO7-8118X (BPM, WNB) and NASA LTSA
grant NAG5-13035 (WNB).

\begin{table}
\begin{center}
\caption{X-ray spectral fitting}
\begin{tabular}{lcccc}
\hline\hline
3C 433 model & $\Gamma$ & $N_{\rm H}$ (10$^{22}$ cm$^{-2}$) & Comments & cstat/d.o.f. \\
\hline
Single-component power-law & & & \\
with partial-covering absorption & 1.44$^{+0.13}_{-0.12}$ & 8.64$^{+0.75}_{-0.64}$ & $f_{\rm c}=0.993^{+0.003}_{-0.002}$ & 596/522 \\
Absorbed power-law & & & \\
with unabsorbed power-law & 1.39$^{+0.22}_{-0.21}$ & 8.25$^{+1.09}_{-0.93}$ & $\Gamma_{\rm unabs} = 2.18^{+0.80}_{-0.83}$ & 594/521\\
Absorbed power-law & & & \\
with unabsorbed thermal & 1.41$^{+0.19}_{-0.24}$ & 8.26$^{+0.94}_{-1.04}$ & $kT = 1.17^{+14.22}_{-0.62}$ & 594/521\\
\hline
4C 65.15 model & $\Gamma$ & $N_{\rm H}$ (10$^{21}$ cm$^{-2}$) & Comments & $\chi^{2}$/d.o.f. \\
\hline
Unabsorbed power-law & 1.89$^{+0.07}_{-0.07}$ & & & 56/82 \\
Absorbed power-law & 1.97$^{+0.11}_{-0.05}$ & 1.31$^{+1.43}_{-1.25}$ & $F$-test p=0.035 & 53/81 \\
\hline
\end{tabular}
\end{center}
\end{table}

\clearpage

\begin{figure}
\includegraphics[scale=0.73]{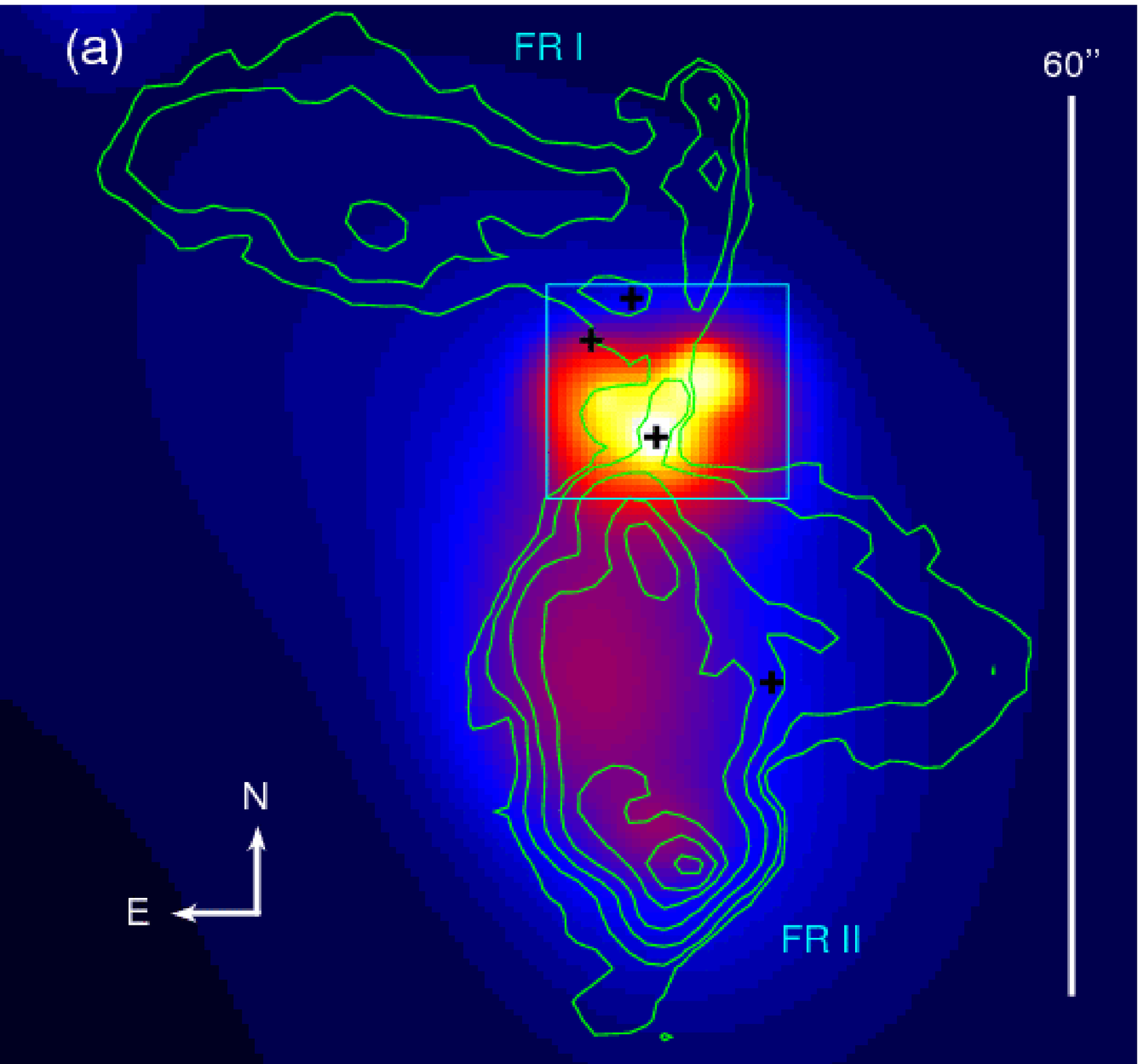}
\includegraphics[scale=0.36]{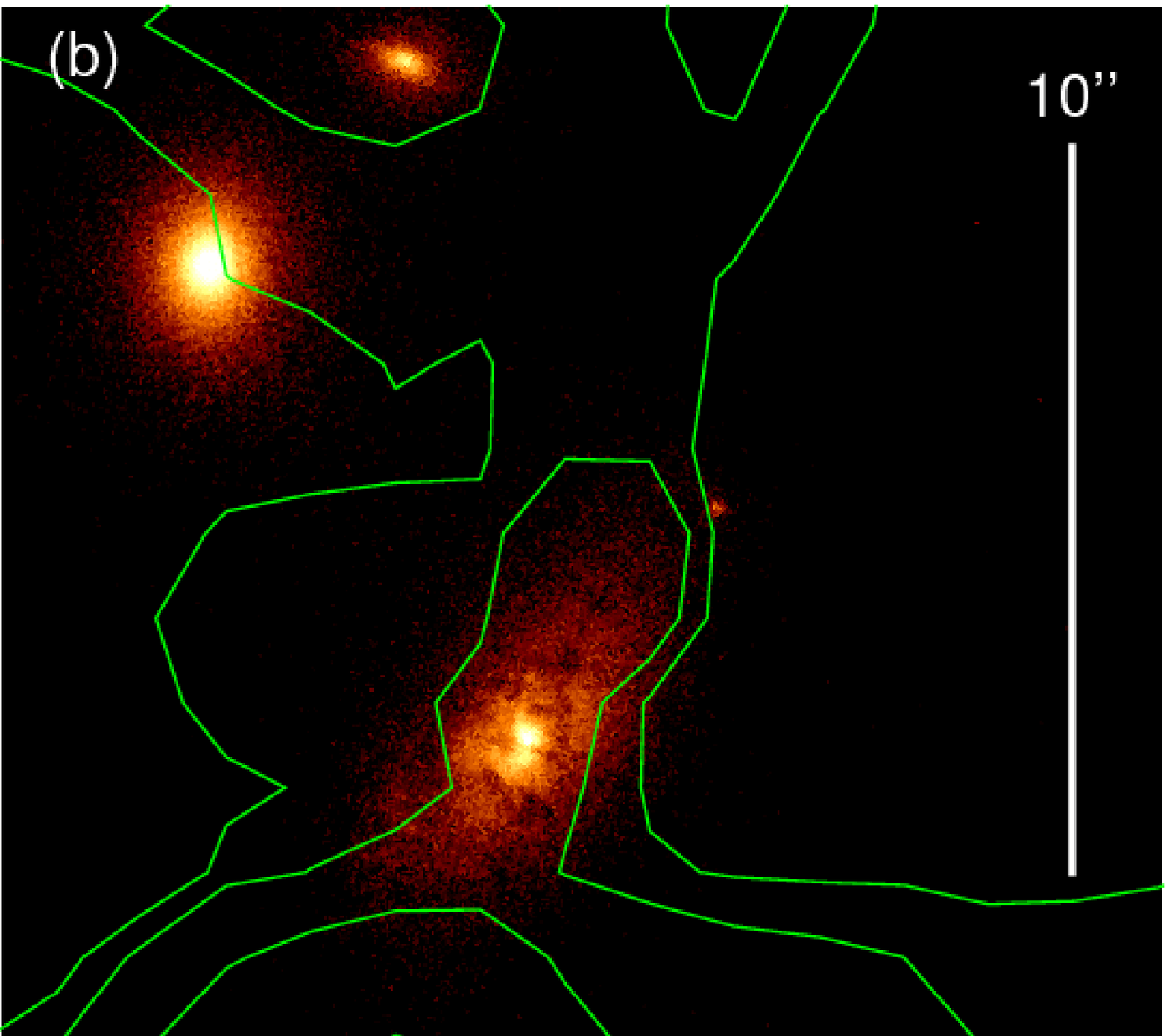}
\includegraphics[scale=0.36]{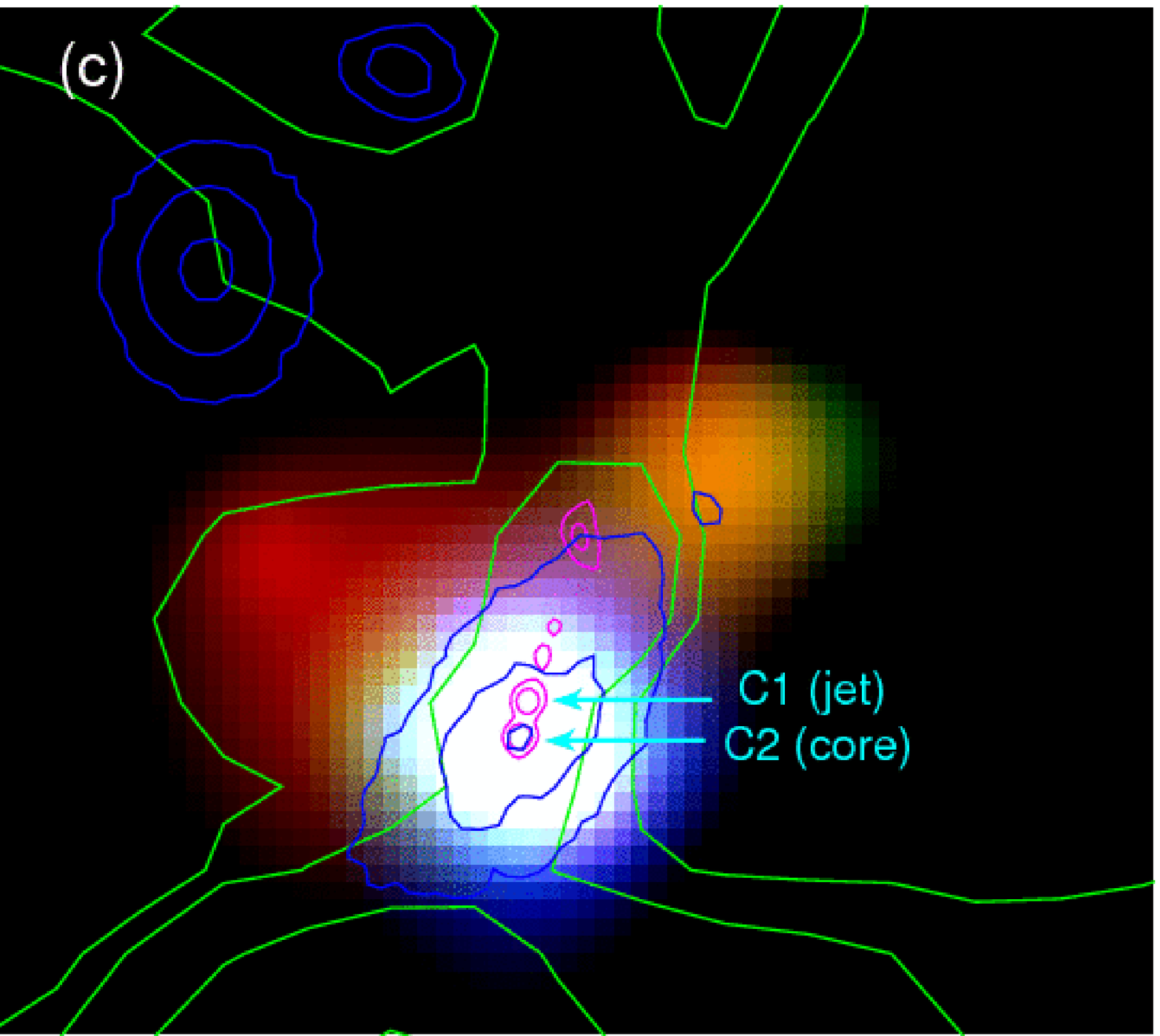} 

\figcaption{\small (a) Adaptively smoothed 0.5--2 keV {\it Chandra} image of
3C~433 overlaid with 1.5~GHz radio contours at levels of 0.7, 2, 7,
20, 40, 55, 70 mJy beam$^{-1}$. Galaxies are marked with crosses. The
square region shows the coverage of the lower panels in this
figure. (b) {\it HST} WFPC2 image of 3C~433; note the dust in the host
galaxy. (c) Smoothed \hbox{X-ray} image color coded by energy band (red is
0.3--1 keV, green is 1--2 keV, and blue is 2--8~keV) with radio and
optical contours overlaid. The magenta contours are high-resolution
8.3 GHz VLA data.}
\end{figure}

\begin{figure}
\includegraphics[scale=0.8]{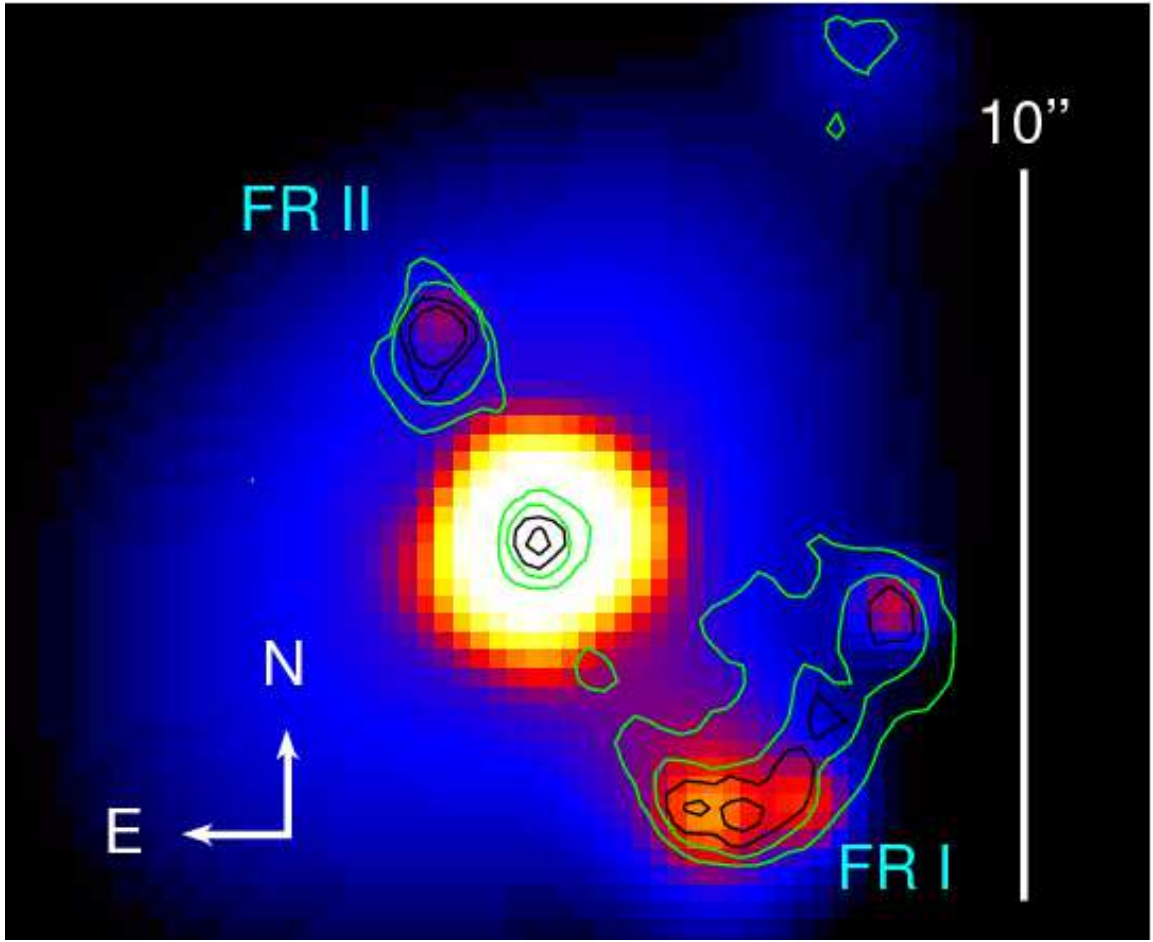} 
\figcaption{\small Adaptively smoothed 0.5--8 keV {\it Chandra} image of
4C~65.15 overlaid with 4.9 GHz radio contours (green) at levels of 0.2
and 1 mJy beam$^{-1}$ and 8.5 GHz contours (black) at levels of 0.5
and 2 mJy beam$^{-1}$. The 4.9 GHz contours (from an image with lower
resolution than that at 8.5 GHz) illustrate the curvature and low
surface brightness expansion of the jet, while the higher frequency
radio data show the knot structure at the bend. There are $\sim$1600
counts in the core and $\sim$30 counts in the extended \hbox{X-ray} feature
at the bend in the FR~I jet. There are $\sim$5 counts near both the
tail of the FR~I jet and the FR~II lobe.}
\end{figure}

\begin{figure}
\includegraphics[scale=0.56,angle=270]{f3a.eps}
\includegraphics[scale=0.69]{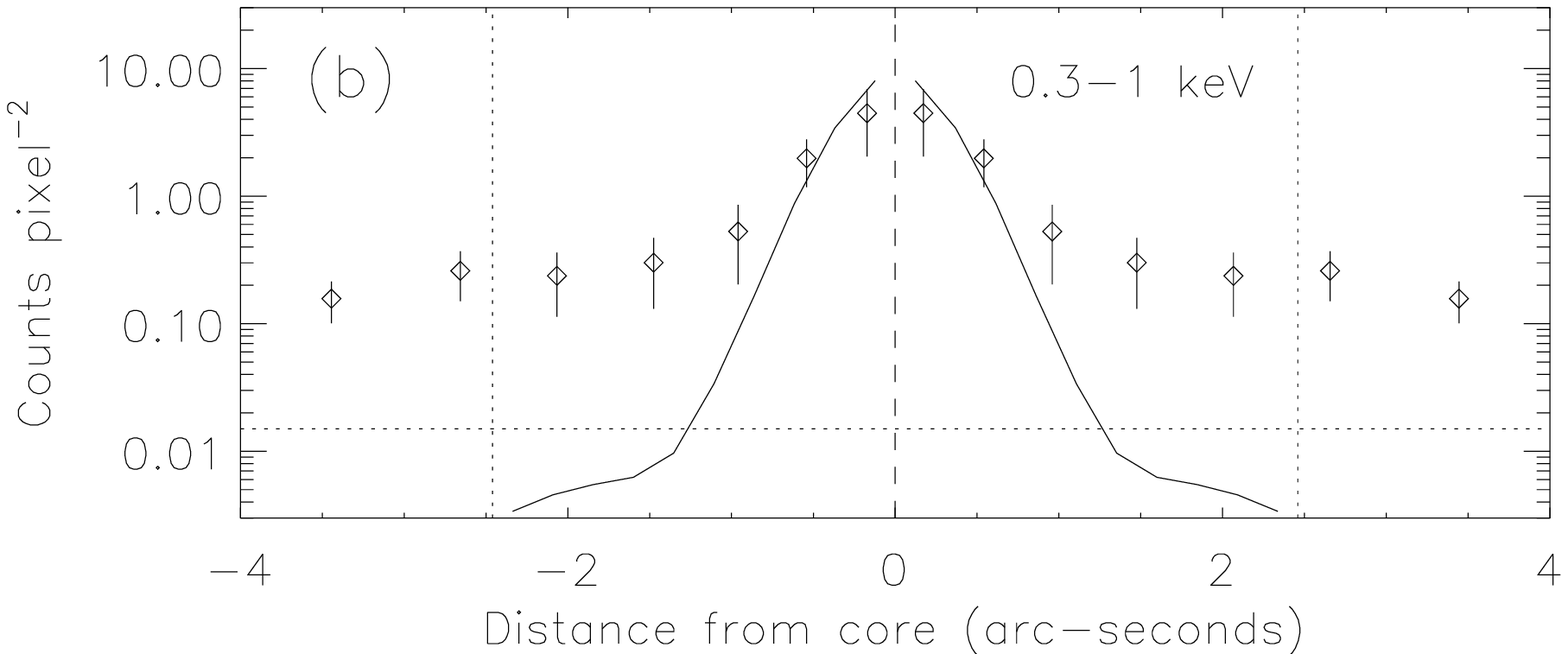} \figcaption{\small (a) {\it
Chandra} 0.3--8 keV spectrum of 3C~433, showing substantial absorption
below 2~keV. For plotting purposes the data were rebinned to have a
minimum significance of 10$\sigma$, with a maximum of 10 bins
combined. The model shown is a double power-law fit with one component
possessing no intrinsic absorption (with $\Gamma$ = 2.2) and the other
absorbed by a neutral column of $N_{\rm H} \simeq
8.3\times10^{22}$~cm$^{-2}$ (with $\Gamma$ = 1.4); this is the second
model in Table~1. The unabsorbed component can be equally well fit
with a thermal bremsstrahlung model with $kT$ = 1.2 keV. (b) 0.3--1
keV radial profile of the 3C~433 core region, constructed from
circular annuli and plotted with a mirrored negative axis for ease of
viewing. The solid line shows a scaled 1~keV point spread function
generated with MARX. The spectrum shown in (a) was extracted from a
circular region $\simeq$2.5$''$ in radius (denoted with vertical
dotted lines); the soft unabsorbed spectral component contains roughly
equal contributions from the diffuse emission and the core. The
diffuse emission persists to $\sim$5--6$''$; the background level is
indicated with a horizontal dotted line.}
\end{figure}

\begin{figure}
\includegraphics[scale=0.8]{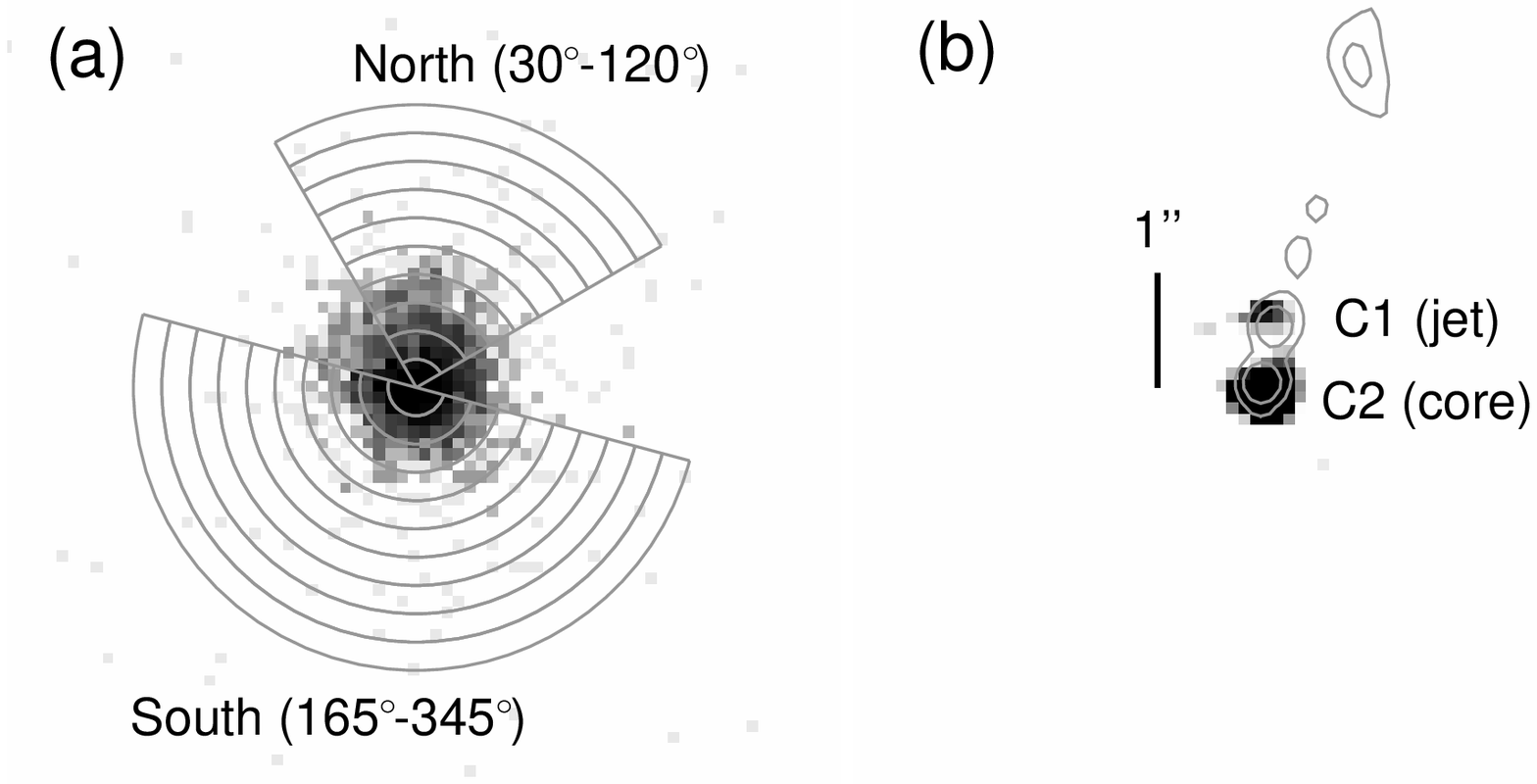}
\includegraphics[scale=0.8]{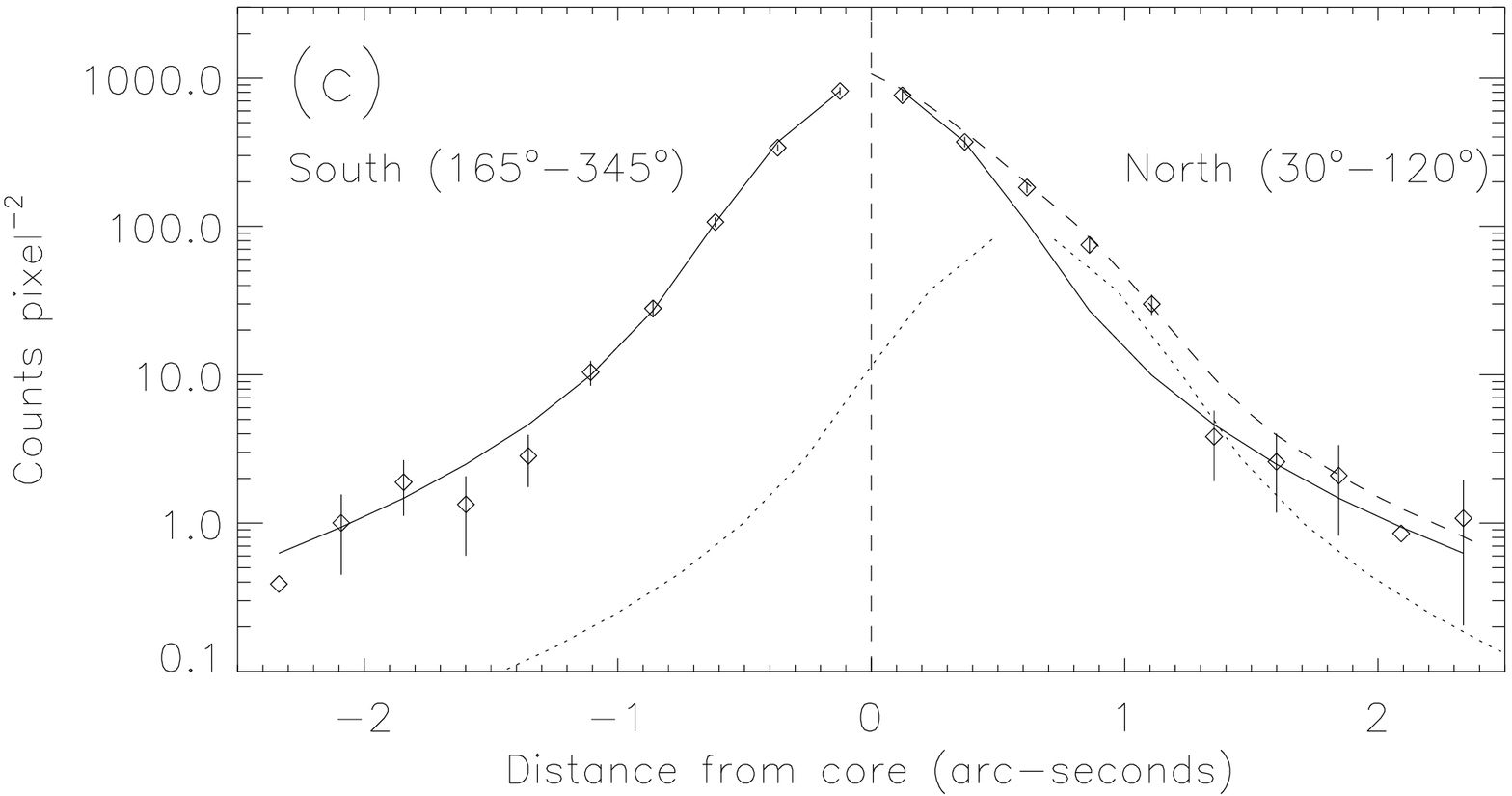} \figcaption{\small (a) {\it
Chandra} 2--6 keV image of the 3C~433 core with pipeline pixel
randomization removed and rebinned to 0.1$''$ pixels. The extraction
regions used to produce the bottom panel in the figure are shown. (b)
A maximum-likelihood reconstruction of the 2--6~keV image overlaid
with 8.3 GHz VLA contours. (c) A radial profile of the 3C~433
core. The solid line shows a scaled 4~keV point spread function
generated with MARX. There is excess \hbox{X-ray} emission to the north,
presumably associated with the northern component of the radio
``double core'' noted by Black et al.~(1992). The dotted line shows a
4~keV point spread function scaled to 10\% of the core and offset by
0.6$''$, and the dashed line shows the superposition of the two point
spread functions.}
\end{figure}

\begin{figure}
\includegraphics[scale=0.7,angle=270]{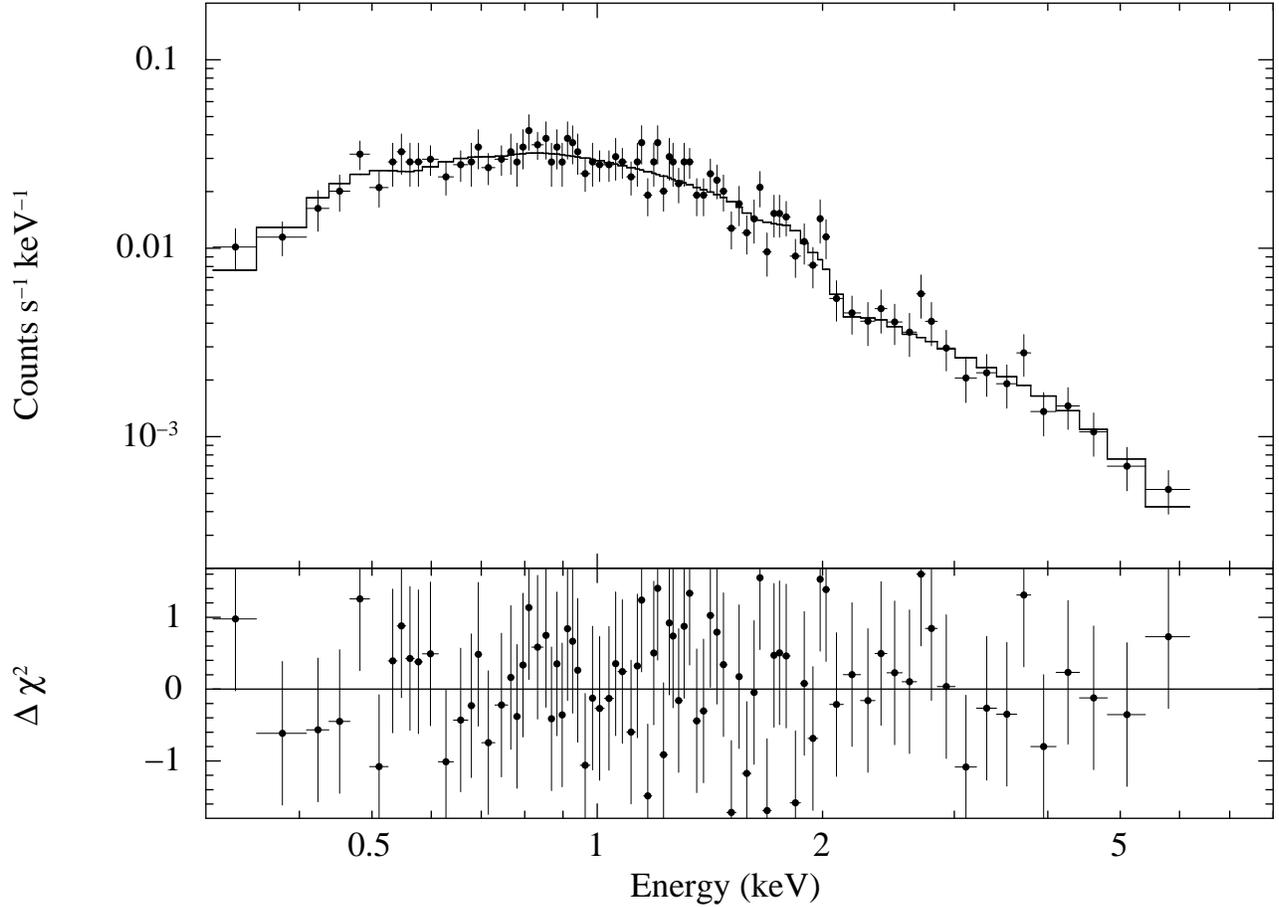}
\figcaption{\small {\it Chandra} 0.3--8 keV spectrum of 4C~65.15, fit with a
power-law model with $\Gamma = 2.0$ and mild intrinsic absorption of
$N_{\rm H} = 1.3\times10^{21}$ cm$^{-2}$. No significant iron emission
is detected. The residuals in the bottom panel are in units of sigma
with error bars of size unity.}
\end{figure}

\begin{figure}
\includegraphics[scale=0.9]{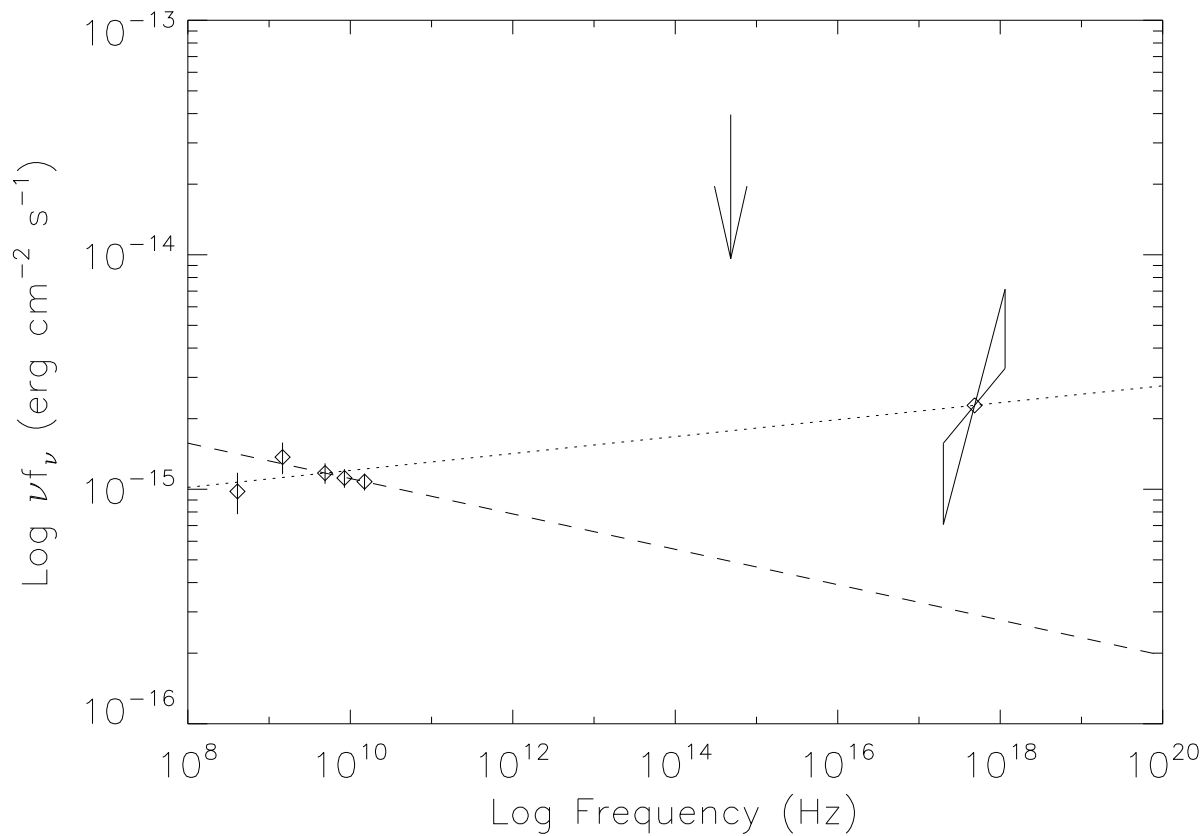}
\figcaption{\small Spectral energy distribution for the jet in 4C 65.15. The
radio fluxes are for the resolved feature at the bend of the jet (see
Figure 2), the optical limit is from the SDSS \hbox{$r$-band} image,
and the \hbox{X-ray} flux and spectral index confidence range are from
fitting the {\it Chandra} data. The dotted line indicates the
${\alpha}_{\rm rx}$ power law calculated between 5~GHz and 2~keV and
the dashed line indicates ${\alpha}_{\rm r}$ calculated between 5~GHz
and 15~GHz. It appears likely that the \hbox{X-ray} data cannot be explained
as arising from the same synchrotron component which generates the
radio flux; deeper optical imaging could confirm this hypothesis.}
\end{figure}

\begin{figure}
\includegraphics[scale=0.9]{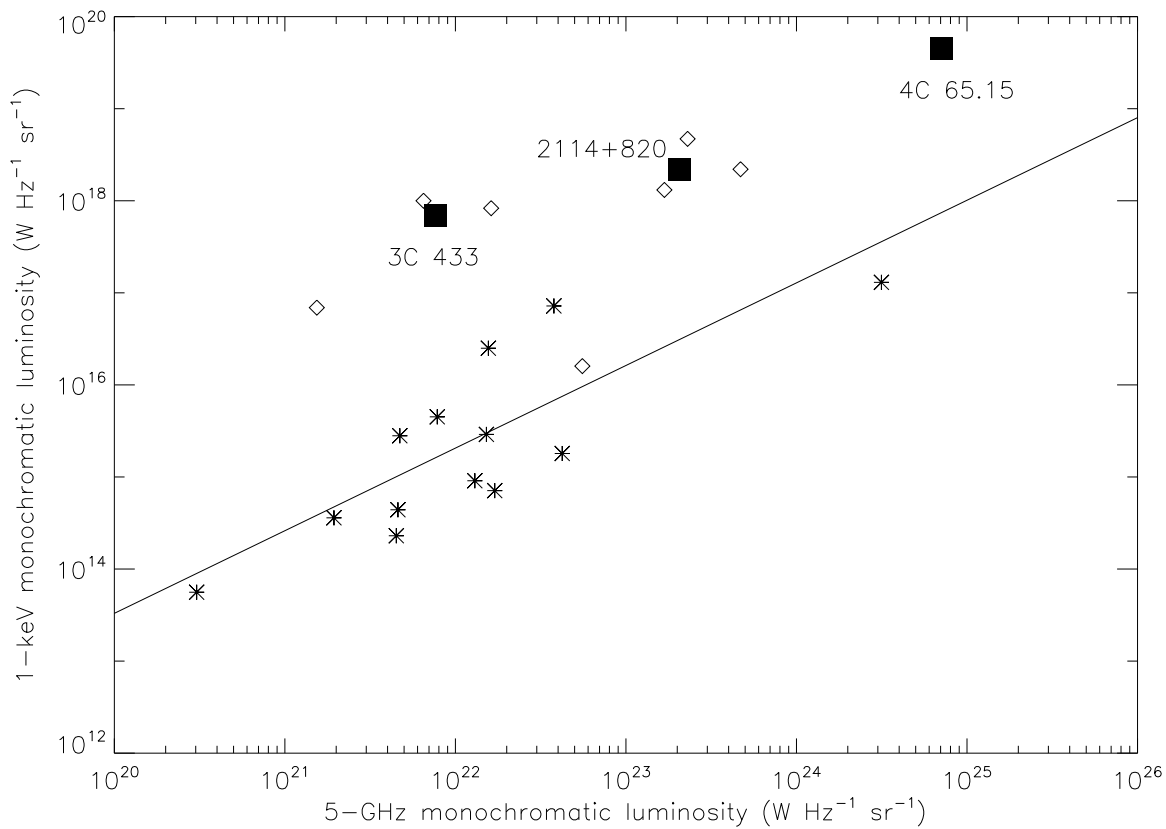} \figcaption{\small Plot of
(unabsorbed) \hbox{X-ray} luminosity versus core radio luminosity with
data from Evans et al.~(2006). Asterisks denote FR~I sources, diamonds
are FR~II sources, and the solid line shows the best fit to the FR~I
luminosity correlation. The low-excitation FR~II 3C~388 and the
absorbed FR~I Cen~A have been omitted. 3C~433 and 4C~65.15 are plotted
as squares; their \hbox{X-ray}/radio properties match those of the
FR~IIs. Also shown is 2114+820, a broad-line object with an FR~I
morphology.}
\end{figure}

\end{document}